\documentclass{jnmp}

\usepackage{amsmath}
\begin{document}
\setcounter{page}{211}
\renewcommand{\evenhead}{C~G\'eronimi, P~G~L~Leach and M~R~Feix}
\renewcommand{\oddhead}{First Integrals and Parametric Solutions}

\thispagestyle{empty}


\FirstPageHead{8}{2}{2001}
{\pageref{geronimi-firstpage}--\pageref{geronimi-lastpage}}{Letter}

\copyrightnote{2001}{C~G\'eronimi, P~G~L~Leach and M~R~Feix}

\Name{First Integrals and Parametric Solutions\\
for Equations  Integrable Through Lie Symmetries}\label{geronimi-firstpage}

\Author{C~G\'ERONIMI~$^{\dag^1}$, P~G~L~LEACH~$^{\dag^2}{}^{\dag^3}$
and M~R~FEIX~$^{\dag^4}$}

\Address{$^{\dag^1}$~MAPMO/UMR 6628, Universit\'e d'Orl\'eans,
D\'epartement de Math\'ematiques\\
$\phantom{^{\dag^1}}$~BP 6759, 45067 Orl\'eans cedex 2, France\\
$\phantom{^{\dag^1}}$~E-mail: geronimi@labomath.univ-orleans.fr\\[2mm]
$^{\dag^2}$~Permanent address: School of Mathematical and Statistical Sciences\\
$\phantom{^{\dag^1}}$~University of Natal, Durban 4041, Republic of South Africa\\[2mm]
$^{\dag^3}$~Member of the Centre for Theoretical and Computational Chemistry\\
$\phantom{^{\dag^1}}$~University of Natal, Durban\\[2mm]
$^{\dag^4}$~Subatech, \'Ecole des mines de Nantes, B.P. 20722, 4 rue Alfred Kastler\\
$\phantom{^{\dag^1}}$~44307 Nantes cedex 3, France}

\Date{Received October 10, 2000; Accepted  March 3, 2001}

\begin{abstract}
\noindent
We present here the explicit parametric solutions of second order
differential equations
invariant under
time translation and rescaling and  third order differential equations
invariant under time translation
and the two homogeneity symmetries.  The computation of first integrals
gives in the most general case, the parametric form
 of the general solution. For some  polynomial functions we obtain a time
parametrisation quadrature which can be
solved in terms
of ``known'' functions.
\end{abstract}

\section{Introduction }

Many works deal with the problem of integrability  of ordinary differential
equations.
Although  the concept of ``integrability'' is delicate, at least four
``routes'' contribute
 to explore it. The first one consists
of  expressing the  general solution of an ordinary differential equation in
 terms of  ``known'' functions; the second deals with the Painlev\'e analysis
 of the equation's singularities; the third uses the Lie symmetry method and
 the last one is  characterized by the existence of invariants, called first
integrals when the independent variable does not explicitly appear.
 Usually a $ N^{\rm th}$ order {\it autonomous}  ordinary differential equation is said to be
``integrable'' if it possesses $N-1$ functionally independent first integrals.
In this paper we  consider the class of ordinary differential equations invariant under
time transla\-tion~($G_1$) and rescaling ($G_2$),  which frequently occur
in  the modelling of natural phenomena.  We write the
generators as
\begin{equation}
G_1 = \p_t, \qquad G_2 = -qt\p_t+x\p_x \qquad (q\in {\mathbb Z}), \label{e1}
\end{equation}
where $t$ is the independent variable and $x$ the dependent variable.
The symmetries $G_1$ and~$G_2$ constitute a representation
of Lie's Type III two-dimensional algebra~\cite{geronimi:lie1}.
The general form of the second order ordinary differential equation
invariant under the action of the two symmetries  $G_1$ and $G_2$  is
\begin{equation}
\ddot x+x^{2q+1}f(\xi)=0,\qquad \xi=\frac{\dot x}{x^{q+1}}\qquad
 ( q \in {\mathbb Z}), \label{geronimi:e2}
\end{equation}
where  $\ddot x$ and $\dot x$ represent respectively ${\rm d}^2 x/{\rm d} t^2$
and  ${\rm d} x/{\rm d} t$.

We also treat the class of third order equations possessing
the three symmetries associated with the generators
\begin{equation}
G_1=\p_t,\qquad G_{21}=t\p_t, \qquad G_{22}=x\p_x.
\label{geronimi:e3}
\end{equation}
The second and third of these, $G_{21}$ and $G_{22}$, are called homogeneity
symmetries~\cite{geronimi:gkp}.
 The general form of a third order ordinary differential equation possessing
the symmetries~(\ref{geronimi:e3})~is
\begin{equation}
\dddot x +\frac{\dot x^3}{x^2}F(\rho)=0,
\qquad \rho=\frac{x\ddot x}{\dot x^2} \cdot \label{geronimi:e4}
\end{equation}
In (\ref{geronimi:e4}) $F$ is an arbitrary function of its argument and
$\rho$ is the second order differential invariant common to $G_1$, $G_{21}$
and
$G_{22}$. Equation  (\ref{geronimi:e4}) is
representative of the class of third order ordinary differential equations invariant under the three-dimensional
algebra with the single nonzero Lie Bracket, $[G_1,G_2] = G_1$, denoted by
$A_1\oplus A_2$ in the Mubarakzyanov classification scheme
\cite{geronimi:mub1,geronimi:mub2,geronimi:mub3}.
We note that the two-dimensional algebra $A_2$ is of
Lie's Type IV since now $G_{21} = tG_1$. We also note that (\ref{geronimi:e2}) is
 naturally connected to (\ref{geronimi:e4}) by the
 Riccati transformation $x=(1+2/q)^{1/q} (\dot{u}/u)^{ 1/q}$.

This paper has two practical goals, {\it viz} firstly to obtain explicit expressions
for the first integrals  needed to declare (\ref{geronimi:e2}) and (\ref{geronimi:e4}) integrable and
secondly
to use these first integrals to push the analytic computations as far as
possible
in order to obtain parametric solutions taking the form of quadratures which
 can be performed in closed form when the functions $f(\xi)$ (in (\ref{geronimi:e2})) and
 $F(\rho)$ (in (\ref{geronimi:e4})) are polynomials.

The paper is divided as follows. In the second section we treat the case
of the second order ordinary differential equation (\ref{geronimi:e2}).
In the third section we show  how to obtain many first
integrals (with of course only two independent) for the third order
ordinary differential equation (\ref{geronimi:e4}) and
how from these first  integrals we deduce the parametric form
of the global solution.
For functions $f$ and $F$ polynomials of at most third degree we present an
explicit solution
of the time parametrisation quadrature. In the fourth section we present our
conclusions.

\section{Parametric solution for a class\\
 of Sordinary differential equation}
The important result is the following reduced form of (\ref{geronimi:e2})
\cite[equation (8)]{geronimi:ref1}
\begin{equation}
\frac{ {\rm d} x}{x}+\frac{\xi {\rm d} \xi}{(q+1) \xi^2+f(\xi)}=0, \label{geronimi:e5}
\end{equation}
which provides the first integral of (\ref{geronimi:e2}).
We  write
\begin{equation}
D(\xi)=f(\xi)+(q+1)\xi^2 \label{geronimi:s1}
\end{equation}
and consequently
\begin{equation}
x=x_0 \exp \left[ -\int_{\xi_0}^{\xi} \frac{u  {\rm d} u}{ D(u)} \right] \label{geronimi:s2}
\end{equation}
In (\ref{geronimi:s2}) the values $x_0$ and $\xi_0$ refer
 to the values of $x$ and $\xi$
 at time $t=0$ which can always be taken as initial time because of the
symmetry
 $G_1$. The parametric solution  of $\dot x$ which is needed to compute the time
scale with $ {\rm d} t= {\rm d} x/\dot x$ is given by $\dot x=\xi x^{q+1}$.
Taking (\ref{geronimi:s2}) into account
\begin{equation}
\dot x= x_0^{q+1} \xi \exp \left[ -(q+1)\int_{\xi_0}^{\xi} \frac{u  {\rm d} u}{
D(u)} \right]. \label{geronimi:s3}
\end{equation}
For polynomial $f(\xi)$ the integrations in (\ref{geronimi:s2}) and (\ref{geronimi:s3}) can be
performed in closed form and for polynomials of degree two the integral  giving $t(\xi)$
can be  expressed in terms of hypergeometric functions.

\section{First integrals and parametric solution for a class\\ of third order
ordinary differential equations}
It was shown \cite[equation (6)]{geronimi:ref1} that a third order
ordinary differential equation invariant under time
translation and
 rescaling symmetries can be written as
\begin{equation}
\dddot x+x^{3q+1}f(\xi, \eta)=0,\qquad \xi=\frac{\dot x}{x^{q+1}},\qquad
\eta=\frac{\ddot x}{x^{2q+1}}, \label{geronimi:e12}
\end{equation}
where $f$ is an arbitrary function of its arguments.
The connection between the two forms as given  by (\ref{geronimi:e12}) and (\ref{geronimi:e4})
(i.e. the relation between $f$ and $F$) is easy to establish. Firstly the
argument $\rho$  is equal to $\eta/ \xi^2$
for all $q$. The identification of
 the two terms in front of $f$ and $F$ gives
\begin{equation}
f(\xi,\eta)=\xi^3 F\left(\frac{\eta}{\xi^2}\right). \label{geronimi:e13}
\end{equation}
 Moreover it was  shown \cite[equation (89)]{geronimi:ref1} that
by applying the two symmetries $G_1$ and
$G_2$ to  (\ref{geronimi:e12}) we obtain
\begin{equation}
\frac{{\rm d } \eta }{{\rm d} \xi
}=\frac{f(\xi,\eta)+(2q+1)\xi\eta}{(q+1)\xi^2-\eta}. \label{geronimi:e14}
\end{equation}
We introduce (\ref{geronimi:e13}) in (\ref{geronimi:e14}), change the variable $\xi$
to $ z= \frac{1}{2} \xi^2$ and reformulate (\ref{geronimi:e14}) with $z$ and
 $\rho$ to obtain separation of variables and the following
first integral
\begin{equation}
 \frac{{\rm d} z}{ z}=\frac{2 (q+1 -\rho)}{F(\rho) +2\rho^2-\rho} {\rm d} \rho.
\label{geronimi:e15}
\end{equation}
Consequently we can replace (\ref{geronimi:e4}) by (\ref{geronimi:e15}) where now $q$  can be
arbitrarily chosen.
Now $q$ being chosen arbitrarily we obtain as many first integrals as we
want, but, of course, only two  will be independent. For polynomial $F(\rho)$,
$q$ can be chosen to exhibit the simplest possible form for these first
integrals.
In order to illustrate our procedure  we consider two  polynomial functions
$F(\rho)$.
Taking first a  quadratic  polynomial we write
 \begin{equation}
F(\rho)+2 \rho ^2-\rho=\gamma (\rho-\alpha_1)(\rho-\alpha_2)\qquad
( \forall \; \gamma,\alpha_1,\alpha_2 \in {\mathbb R} \ {\rm and} \ \alpha_1 \not=
\alpha_2) \label{geronimi:e16}
\end{equation}
(the case $\alpha_1=\alpha_2$ is similar).
Now we take in (\ref{geronimi:e15})
 $q_1=\alpha_1-1$. Introducing (\ref{geronimi:e16}) in (\ref{geronimi:e15})  we obtain after a
simple integration
 the first integral
\begin{equation}
I_1= z_1^{\gamma/2}(\rho-\alpha_2) = \left(\frac{ \dot x}{x^{ \alpha_1}}\right)^{\gamma}
(\rho-\alpha_2). \label{geronimi:e17}
\end{equation}
To find a second simple first integral, we take $q_2=\alpha_2-1$
 and obtain
\begin{equation}
 I_2= z_2^{\gamma/2}(\rho-\alpha_1) = \left(\frac{ \dot x}{x^{ \alpha_2}}\right)^{\gamma}
(\rho-\alpha_1). \label{geronimi:e18}
\end{equation}
 So far  we have obtained two rather simple first integrals and  the ordinary
differential equation is
 now integrable. Moreover the time parametrisation of the trajectory is
obtained by the elimination
of $\rho$ between $I_1$ (\ref{geronimi:e17}) and $I_2$ (\ref{geronimi:e18}).  This gives
\begin{equation}
\dot x =\left( \frac{I_2 x^{\alpha_2 \gamma} -I_1 x^{\alpha_1
\gamma}}{\alpha_2-\alpha_1}\right)^{1/\gamma}
 \label{geronimi:e19}
\end{equation}
 and implies the time parametrisation quadrature
\begin{equation}
t=  (\alpha_2-\alpha_1)^{1/\gamma}
\int \frac{{\rm d} x}{\left( I_2 x^{\alpha_2 \gamma} -I_1 x^{\alpha_1
\gamma}\right)^{1/\gamma}  }\cdot \label{geronimi:q1}
\end{equation}
Not surprisingly the performance of the quadrature in (\ref{geronimi:q1}) is generally
not possible in closed form.
We turn now to an $F(\rho)$ which is a
 third degree polynomial in $\rho$. We write
 \begin{equation}
\arraycolsep=0em
\begin{array}{l}
F(\rho)+2 \rho ^2-\rho=\gamma (\rho-\alpha_1)(\rho-\alpha_2)(\rho-\alpha_3),
\vspace{1mm}\\
 ( \forall \; \gamma,\alpha_i \in {\mathbb R} \ {\rm and} \ \ \forall \; i \not= j, \
\alpha_i \not= \alpha_j).
\end{array} \label{geronimi:ep1}
\end{equation}
(If at least two roots $\alpha_i$ are equal, the problem is quite similar).
Taking respectively $q_i =\alpha_i-1$,
$(i=1,2,3)$, in (\ref{geronimi:e15}) and after a
direct integration we
find the first integrals
\begin{equation}\arraycolsep=0em
\begin{array}{l}
\displaystyle \frac{\dot x}{x^{\alpha_1}}^{\gamma (\alpha_2-\alpha_3)}
 (\rho -\alpha_2) (\rho -\alpha_3)^{-1}=I_1
\vspace{3mm}\\
\displaystyle  \frac{\dot x}{x^{\alpha_2}}^{\gamma (\alpha_3-\alpha_1)}
 (\rho -\alpha_3) (\rho -\alpha_1)^{-1}=I_2
\vspace{3mm}\\
\displaystyle \frac{\dot x}{x^{\alpha_3}}^{\gamma (\alpha_1-\alpha_2)}
 (\rho -\alpha_1) (\rho -\alpha_2)^{-1}=I_3. \label{geronimi:ep2}
\end{array}
\end{equation}
The relation
\begin{equation}
I_1 I_2I_3=1 \label{geronimi:ep3}
\end{equation}
 proves that the three first integrals are not independent. Thus any pair of
(\ref{geronimi:ep2})
 provides the two independent first integrals needed to declare the
ordinary differential equation
integrable.

We return now to the general problem. From (\ref{geronimi:e15}) writing
$q+1=\alpha_i$,  $i=1,2$, we obtain
\begin{equation}
\frac{\dot x^2}{x^{2 \alpha_i}}=\frac{\dot x_0^2}{x_0^{2 \alpha_i}}\exp
\left[
\int_{\rho_0}^{\rho}\frac{2(\alpha_i-\rho)}{D(\rho)} {\rm d} \rho
\right],\qquad i=1,2,\label{geronimi:333}
\end{equation}
where $D(\rho)=F(\rho)+2\rho^2-\rho$. Firstly we eliminate $\dot x$ between the two
equations (\ref{geronimi:333}) to give
\begin{equation}
x^{2(\alpha_2-\alpha_1)}=x_0^{2(\alpha_2-\alpha_1)}
\exp
\left[
\int_{\rho_0}^{\rho}\frac{2(\alpha_1-\alpha_2)}{D(\rho)} {\rm d} \rho
\right]
\end{equation}
and consequently
\begin{equation}
x=x_0\exp \left[ -\int_{\rho_0}^{\rho} \frac{ {\rm d} u}{D(u) }
\right]. \label{geronimi:b1}
\end{equation}
Then we eliminate $x$ to obtain $\dot x$
\begin{equation}
\dot x=\dot x_0 \exp
\left[
-\int_{\rho_0}^{\rho} \frac{u {\rm d} u}{D(u) }.
\right] \label{geronimi:b2}
\end{equation}
Relations (\ref{geronimi:b1}) and (\ref{geronimi:b2}) give the parametric solution
(with  $ {\rm d} t = {\rm d} x/
\dot x$).
 Although (\ref{geronimi:b1}) and (\ref{geronimi:b2}) can be obtained directly (computing
$\dot{\rho}$ and
$\dot{\rho} \rho$) the rather simple expression for the two first integrals
is obtained
 by selecting $q_i +1=\alpha_i$, $i=1,2$, in (\ref{geronimi:e15}), $\alpha_1$ and
$\alpha_2$
 being two roots of $D(\rho)=0$.  Finally we observe that the computation
of the leading
order term in the Painlev\'e analysis with a singularity in
$(t-t_0)^{-1/q}$ gives the
equation
\begin{equation}
(1+q)(1+2q)+F(1+q)=0.
\end{equation}
If we  write $\alpha=1+q$, this last equation becomes
\begin{equation}
F(\alpha)+2\alpha^2-\alpha=D(\alpha)=0.
\end{equation}
This confirms the asymptotic character of some of these singular solutions
 since we recover the divergence of $x$ in (\ref{geronimi:b1}) for exactly the same
value $q+1 = x \ddot x /\dot x^2$ when $x \sim (t-t_0)^{-1/q}$.

\section{Conclusion}
This work deals with the possibility of exhibiting  explicit first integrals
and   parametric solutions for  second and third order ordinary differential
equations possessing the necessary number of
symmetries
to be formally integrable. Quite naturally the parameters allowing these
simple parametric solutions are the two differential invariants of the
equation, namely~$\xi$ for the second order equation and  $\rho$  for the third order
equation. For functions $f(\xi)$ and
$F(\rho)$ polynomial
 the integrals giving $x$ and $\dot x$ can be performed in closed form and for
quadratic $f(\xi)$
 the integral giving $t$ is obtained in terms of hypergeometric functions.
Finally for pathological  functions $f$ and $F$  our method minimises
 numerical calculations. Moreover for the
 third order equation the main point
is indeed the identity with the two homogeneity symmetries and rescaling
with  an arbitrary parameter $q$ which makes our method systematic. For the
obtention  of first integrals, another aspect of
these results is that they are at the same time connected and complementary
to the Painlev\'e analysis.

\subsection*{Acknowledgements}
PGLL thanks Professor M R Feix and MAPMO, Universit\'e d'Orl\'eans, for
their kind hospitality while this work was undertaken and the National
Research Foundation of South Africa and the University of Natal for their continuing
support.

\label{geronimi-lastpage}

\begin{thebibliography}{99}
\small
\topsep0mm
\partopsep0mm
\parsep0mm
\itemsep0mm

\bibitem{geronimi:lie1}
Lie S, Differentialgleichungen, Chelsea - New York, 1967.
\bibitem{geronimi:gkp}
Flessas G P, Govinder K S and Leach P G L, Characterisation of the Algebraic
Properties of First Integrals of Scalar Ordinary Differential
Equations of Maximal Symmetry,  {\it J.~Math. Anal. Appl.} {\bf 212}
(1997), 349--374.
\bibitem{geronimi:mub1}
Mubarakzyanov G M,  On Solvable Lie Algebras, {\it Izvestia Vysshikh Uchebnykh
Zavendeni\u{\i}, Matematika} {\bf 32} (1963), 114--123.
\bibitem{geronimi:mub2}
Mubarakzyanov G M,  Classification of Real Structures of Five-Dimensional Lie
Algebras, {\it Izvestia Vysshikh Uchebnykh
Zavendeni\u{\i}, Matematika} {\bf 34} (1963), 99--106.
\bibitem{geronimi:mub3}
Mubarakzyanov G M,  Classification of Solvable Six-Dimensional Lie Algebras
with One Nilpotent Base Element, {\it Izvestia Vysshikh Uchebnykh
Zavendeni\u{\i}, Matematika} {\bf 34} (1963), 104--116.
\bibitem{geronimi:ref1}
Feix M R, Geronimi C, Cair\'{o} L, Leach P G L, Lemmer R L and
Bouquet S \'{E}, On the Singularity Analysis of Ordinary Differential Equation
Invariant under Time Translation and Rescaling, {\it J. Phys. A:
Math. Gen.} {\bf 30} (1997), 7437--7461.
\bibitem{geronimi:pms88}
Leach P G L, Feix M R and Bouquet S, Analysis and Solution of a Nonlinear
Second-Order Equation Through Rescaling and Through a Dynamical Point of
View, {\it J.~Math. Phys.} {\bf 29} (1988),  2563--2569.
\bibitem{geronimi:smp91}
Bouquet S \'E, Feix M R  and Leach P G L, Properties of Second-Order
Ordinary Differential Equations
Invariant under Time Translation and Self-Similar Transformation, {\it
J.~Math. Phys.} {\bf 32} (1991), 1480--1490.
\end{thebibliography}
\end{document}